\documentclass{article}

\usepackage{arxiv}

\usepackage[utf8]{inputenc} % allow utf-8 input
\usepackage[T1]{fontenc}    % use 8-bit T1 fonts
\usepackage{hyperref}       % hyperlinks
\usepackage{url}            % simple URL typesetting
\usepackage{booktabs}       % professional-quality tables
\usepackage{amsfonts}       % blackboard math symbols
\usepackage{nicefrac}       % compact symbols for 1/2, etc.
\usepackage{microtype}      % microtypography
\usepackage{lipsum}		% Can be removed after putting your text content
\usepackage{graphicx}
\usepackage{graphicx}
\usepackage{natbib}
\usepackage{doi}

\title{Evaluation of Active Affiliates to the SIS Multidimensional Analysis in R Shiny}

\author{ACeituno-Moya, Nadine\\
Faculty of Statistics and Computer Engineering\\ National University of the Altiplano of Puno, \\
P.O. Box 291, Puno - Perú \\
\texttt{naceituno@est.unap.edu.pe} \\
\and 
Torres-Cruz, Fred \\
Faculty of Statistics and Computer Engineering\\ National University of the Altiplano of Puno,\\
P.O. Box 291, Puno - Perú\\
\texttt{ftorres@unap.edu.pe}
}

\renewcommand{\shorttitle}{\textit{Evaluation of Active Affiliates to the SIS Multidimensional Analysis in R Shiny} }

\date{}
\begin{document}

\maketitle

\begin{abstract}
	This article presents a study that applies multiple linear regression analysis to active affiliates of the Comprehensive Health Insurance (SIS) in Peru. The main objective is to examine the factors that can affect the number of people affiliated with different insurance plans.

    The study emphasizes the significance of multiple linear regression analysis in understanding the factors that influence the affiliates of SIS Comprehensive Health Insurance. Additionally, it demonstrates the value of using interactive tools like RShiny to enhance data analysis by providing a dynamic and participatory experience for researchers and users interested in the topic.

    To facilitate the analysis and visualization of SIS-related data, an interactive application was developed using RShiny. This tool enables the loading, visualization, and analysis of data in a user-friendly and practical manner. By offering an interactive platform, users can effectively explore and comprehend the factors impacting SIS affiliates.

    The analysis results reveal that the selected variables exert a significant positive influence on the total number of affiliates. This suggests that the specific plan examined in this study has a favorable impact on the enrollment of individuals in SIS. Furthermore, the data exhibits a linear trend, supporting the appropriateness of employing a linear regression model to describe this relationship.\\

\end{abstract}

% keywords can be removed

\section{Introduction}

    Comprehensive health insurance (SIS) is a public institution in Peru that aims to provide access to health services for low-income people. In this context, it is important to understand how different factors can influence the number of SIS affiliates in different regions of the country. Multiple linear regression analysis is a statistical tool that allows you to explore relationships between multiple predictor variables and a response variable, which can help you understand how these factors affect membership in different regions.\\

    In this article, multiple linear regression analysis will be applied to active SIS affiliates in order to understand how different factors can affect the number of affiliates in different regions of Peru. To this end, the study entitled "Sense Impact of Comprehensive Health Insurance (SIS) from Beneficiaries and Providers"Tumbes Region--Peru" \cite{solis2018impacto} will be used as a reference. In addition, other related studies will be taken into account, such as "Access to Medicines in Comprehensive Health Insurance (SIS) Patients with Diabetes Mellitus and/or Arterial Hypertension in Peru" \cite{espinoza2021acceso}.\\

    The article will focus on the use of interactive tools, specifically an application based on RShiny, to improve the visualization of the data and facilitate the analysis of the factors that affect the number of affiliates to the SIS. These tools will allow users to select the regions and range of affiliated individuals they wish to analyze, providing an accessible and hands-on experience when exploring the data\cite{potter2016web}.\\

    The interactive application developed with RShiny will allow you to view the results of the multiple linear regression analysis dynamically. Users will be able to select the predictor variables they wish to analyze, such as socioeconomic level, availability of health services, and geographic accessibility, among others. In addition, they will be able to explore different regions of Peru and observe how these factors influence the number of SIS affiliates in each of them\cite{garcia2022datos}.\\

    The study will be based on data obtained from the National Data Platform (gob.pe) corresponding to the period 2023-5. These data are of vital importance for the analysis since they offer updated and relevant information on active members of the Comprehensive Health Insurance (SIS) in different regions of Peru. By using updated data, we can obtain more accurate and representative results regarding the current situation regarding SIS affiliation\cite{hernandez2015determinantes}.

    The National Data Platform is a reliable and official source that collects detailed information on various aspects of health in Peru, including data related to the SIS. By accessing this platform, it is possible to obtain data broken down by region, which is essential to understanding the variations in the number of SIS affiliates in different parts of the country.

    In summary, this article aims to improve our understanding of how different factors can affect the number of affiliates to the Comprehensive Health Insurance (SIS) in different regions of Peru. Multiple linear regression analyses will be applied, and an interactive application based on RShiny will be used to facilitate data exploration and analysis. The use of these interactive tools will allow a better visualization of the data and provide valuable information on the factors that influence the number of affiliates to the SIS.

\section{MATERIALS}

    The materials used for the realization of this code are the following R libraries:
    \begin{figure}[htp]
        \centering
        \includegraphics[width=0.7\linewidth]{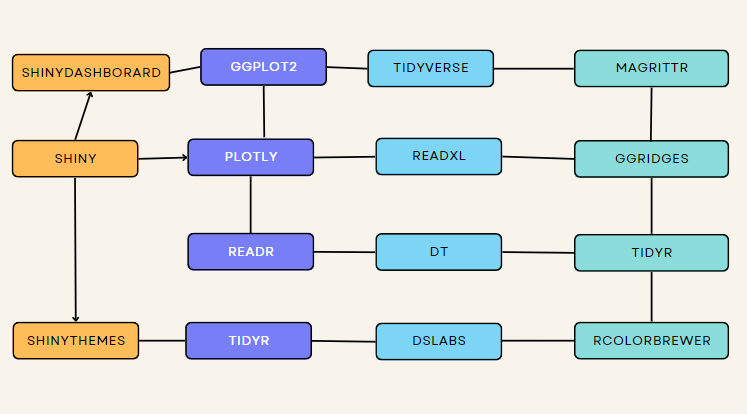}
        \caption{Flowchart: Libraries}
        \label{fig:my_label}
    \end{figure}
    
    These libraries are used to load and manipulate data, create the application's graphical interface, generate interactive graphs, perform statistical calculations, and customize the appearance of the application.

\section{METHODOLOGY} 
    The methodology used in this interactive exploration of active members of comprehensive health insurance (SIS) through RShiny is divided into several key stages.\\
    
    Data collection and preparation: The data of active members of the SIS were obtained, including demographic information, from a secure page, and with constant updating, said data was taken from the last update (2023-5). The data was organized in a suitable format for analysis.\\
    
    Multiple Linear Regression Analysis:
    The simple linear regression model has been studied, where the influence of an explanatory variable X on the values taken by another variable called the dependent variable (Y) was analyzed.\cite{salinas2007modelos}point out that multiple linear regression supposes that more than one variable has influence or is correlated with the value of a third variable. In this case, the variables \textit{insurance plan, inei scope and age} will be used as predictor variables to understand their influence on the total number of affiliates to the SIS.\cite{dagnino2014regresion} mentions that linear regression allows predicting the behavior of a dependent variable from one or more predictor variables. In addition, they highlight the importance of verifying the assumptions of linearity, normality, and homogeneity of variances, as well as performing an analysis of the point cloud and residuals to assess the validity of the model.\\
    
    Development of the RShiny application: An interactive application was created using RShiny that allows users to select the variables of interest for analysis. \cite{san2023aplicaciones}User interface elements such as dropdowns and sliders have been implemented to make it easier to select regions and other relevant variables.\\
    Next, we show you the structure of the application:\\

    \begin{figure}[htp]
	\centering
	\fbox{\includegraphics[width=.8\linewidth]{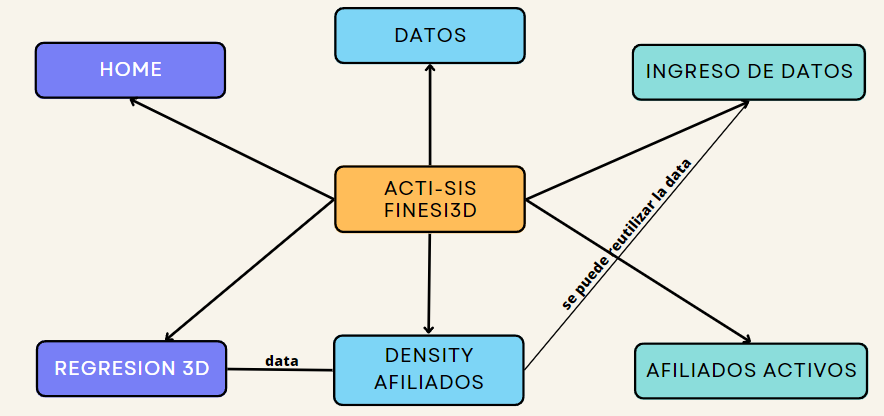}}
	\caption{Flowchart: Application}
    \end{figure}

    This methodology combines data collection, interactive application development in RShiny, data visualization, and multiple linear regression analysis. It provides users with an accessible and practical tool to explore and understand SIS Comprehensive Health Insurance enrollment factors in an interactive and hands-on way.\\
\section{Data verification process}
    First, the data was obtained from the SIS Comprehensive Health Insurance affiliate database ($gob.pe$). Subsequently, we combine the data that is delimited by commas in two apps, R-Studio and Excel.\\
    
    Subsequently, a data cleaning, known as "data cleaning" or "data scrubbing,", was performed on our data. The data type of each variable was analyzed and corrected to ensure its efficiency and manipulability during our analyses\cite{rahm2000data}.\\
    
    Our database initially consisted of 1,048,576 records. We work with this entire population and also carry out a corresponding sampling, obtaining a subset of 9,805 records.\\
    
    After considering these points, we verified the variables and decided to analyze the multiple linear regression of the total active members of the Comprehensive Health Insurance (SIS). To do this, we use the following variables: \textit{"Insurance Plan", "Region", "Age", "National Foreigner," and "INEI Scope." }. \\
\\ \\
    \begin{table}[htp]
	\caption{Variable Dictionary}
	\centering
        \begin{tabular}{ccc}
            \hline
                ATTRIBUTE & DESCRIPTION\\
            \hline
                REGION & Region of residence \\
            \hline
                AGE & Age calculated at the moment \\
                & of the consultation in the Base\\
                & of SIS Data\\
            \hline
                NATIONAL & Resident of Nationality\\
                FOREIGN & Peruvian and resident \\
                &  foreign \\
            \hline
                SCOPE & Geographic Description\\
                INEI & according to INEI\\
            \hline
                INSURANCE PLAN & Type of insurance SIS,\\
                & associated with the list of \\
                & benefits (health benefits)\\
                & that provides your financial coverage\\
                & (SIS Free, SIS For All,\\
                & Independent SIS, SIS NRUS \\
                & and SIS Microenterprises)\\
            \hline
                TOTAL OF & Amount of affiliates according to the \\
                AFFILIATES & combination of variables \\
                & of row \\
            \hline
        \end{tabular}
    \end{table}
\section{MULTIPLE LINEAR REGRESSION}
    \subsection{Sampling}
        Defining the function calculate sample size: Create a function called calculate sample size that takes three arguments: population size (population size), confidence degree (confidence level), and margin of error (margin of error). This function will calculate the sample size needed to get accurate estimates\cite{sanchez2009analisis}.
\\

        Calculation of the z value: The q norm function is used to calculate the z value corresponding to the confidence level. The z value is obtained by dividing the sum of 1 and the confidence level by 2.\\

        Establishment of the maximum ratio: A conservative maximum ratio of 0.5 (p = 0.5) is established. This value is used to calculate the required sample size in the event that there is no prior information on the actual proportion\cite{lopez2004poblacion}.
\\
        Sample size calculation: The formula for the calculation of the random and effective sample size is applied, which is based on the formula for sample proportions.\\
        The following formula is used:

            $$
            n =\frac{z^2*p*q}{{margen error^2}+\frac{z^2*p*q}{poblacion size}}
            $$
        Where:

        n: required sample size\\
        z: z-value for the confidence level\\
        p: maximum conservative ratio\\
        q: complement of the maximum conservative ratio\\
        margin of error: margin of error allowed\\
        population size: population size\\

        Rounding of the sample size: The result obtained in the calculation of the sample size can be a decimal number. To ensure that the sample size is an integer, the ceiling function is used to round up to the nearest integer\cite{torres2006tamano}.

        Return sample size: The sample size calculated as a result of the calculate sample size function is returned.

        Definition of the parameters: The values of the parameters necessary for the calculation of the sample size are defined. These include population size (population size), confidence degree (confidence level), and margin of error (margin of error)\cite{lopez2004poblacion}.

        Sample size calculation: The function calculate sample size is called by passing the parameters defined above to obtain the required sample size.

        Show the calculated sample size: The sample size calculated using the print function is printed on the screen.

    \subsection{Regression}
        Perform Multiple Linear Regression: The lm function is used to fit a multiple linear regression model. The formula specifies the dependent variable \textit{(Total Affiliates)} and the independent or predictor variables \textit{(Insurance  plan, region, age, foreign national, INEI scope)} that will be used to predict the dependent variable\cite{abuin2007regresion}.
        The data argument indicates the data set from which the variables will be obtained.

        Model Fit: The lm function is used to fit the multiple linear regression model. The result is assigned to the variable lm model, which will contain the estimated coefficients and other information related to the model\cite{salinas2007modelos}.
        The formula of the multiple linear regression model used:
            $$
            \textit{Total Afiliados} = \beta_0 + \beta_1*\textit{Plan de Seguro} + \beta_2*\textit{Región} + 
            $$
            $$
            \beta_3*\textit{Edad} + \beta_4*\textit{Nacional Extranjero} + \beta_5*\textit{Ambito INEI} +\varepsilon
            $$

        Where:

        \textit{ Total Affiliates}: is the dependent variable, which represents the number of affiliates.
        \textit{ Insurance Plan, Region, Age, Foreign National, and INEI Scope} are the independent or predictor variables used to predict TOTAL AFFILIATES.

            $$\beta_1,\beta_2,\beta_3,\beta_4$$
        Are the estimated coefficients of the model, which represent the relationship between the predictor variables and the dependent variable\cite{abuin2007regresion}.
            $$\varepsilon $$
        Is the error term, which represents the variation not explained by the model.

        Get the predictions: The predict function is used to get the predictions of the fitted model. The lm model is passed as an argument, and no additional data set is specified, which means that the model fit data will be used to perform the predictions \cite{dagnino2014regresion}.

        Store the predictions: The model's predictions are assigned to the predicted values variable, which will contain the predicted values for the dependent variable \textit{(Total Affiliates)} based on the predictor variables used in the model.
        
\section{THE APP}
    \subsection{Functionality}

        For the development of the application, the powerful tools of  RStudio and RShiny were used. These tools provided an efficient and versatile environment for building the application, allowing for agile deployment and an interactive user interface\cite{san2023aplicaciones}.

        The program provides a graphical user interface (UI) for the analysis and visualization of data related to comprehensive health insurance (SIS). Here is a description of the main functions of the program:
        
        Home: This tab shows general information about comprehensive health insurance, including a brief description and the benefits offered by the insurance. Information is also provided on SIS plans and how to keep SIS active. In addition, links are included to access more information about insurance plans and to verify if a person is affiliated with SIS.\\

        Data: In this tab, users can upload a data file in CSV or XLSX format that contains the information of the active members of the SIS. Once the data is loaded, a table with the data is displayed, and a summary of the data is provided. You can also select a variable and generate distribution plots to visualize the distribution of the data\cite{beeley2016web}.\\

        Data Entry: This tab allows you to upload a data file and perform additional filters. Users can select fields or columns to filter and choose display options such as bar charts or pie charts. Charts are generated based on the user's selections.\\

        Regression Analysis: In this tab, users can upload a data file and perform a multiple linear regression analysis. The dependent variable \textit{(Total Affiliates)} and the predictor variables \textit{"Insurance Plan", "Region", "Age", "National Foreigner" and "INEI Scope" are used. } 3D scatter plots and correlation graphs to analyze the relationship between variables

        Affiliate Density: In this tab, users can upload a data file and view affiliate density on a graph. A variable can be selected to generate a density plot\cite{jimenez1991modelizacion}.

        Active Affiliates: In this tab, users can upload a data file and filter the data by region. A map is displayed with bars and dots representing active affiliates in each region.
 
        \begin{figure}[htp]
            \centering
            \fbox{\includegraphics[width=.6\linewidth]{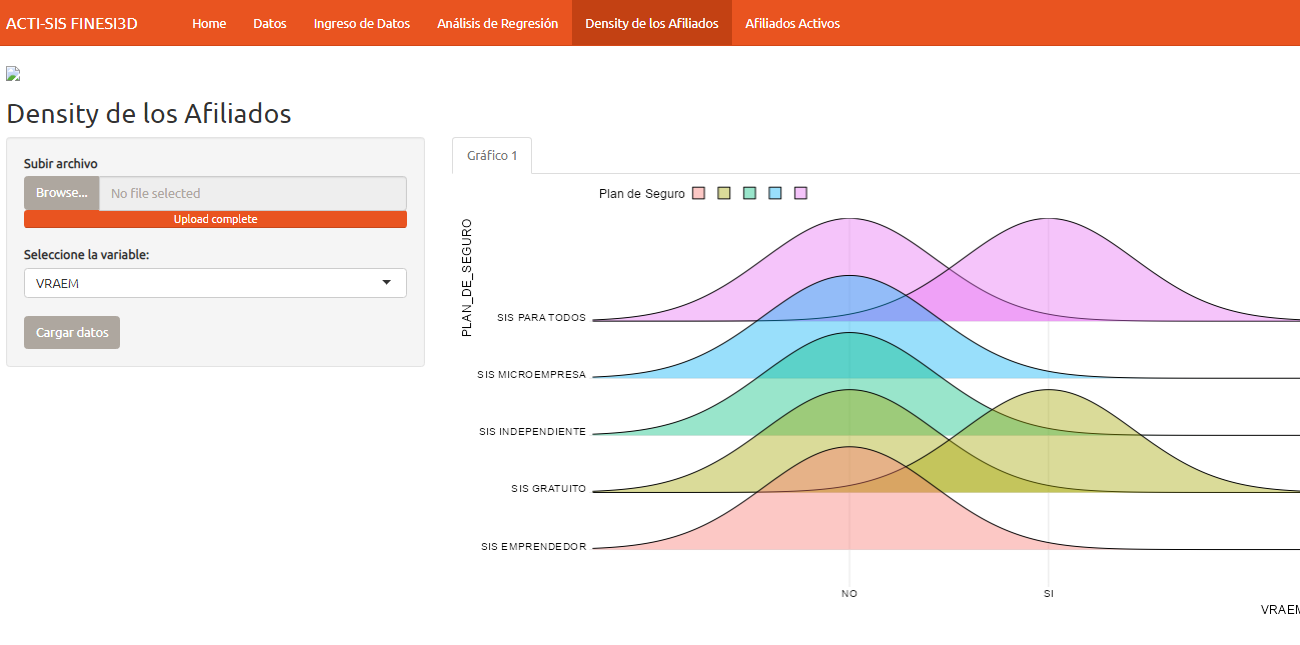}}
                \caption{The Applicative}
        
        \end{figure}

\section{RESULTS}

    We observe the results of the multivariate regression for the active members of the comprehensive health insurance system (SIS).\\
    
    \begin{figure}[htp]
        \centering
        \fbox{\includegraphics[width=.6\linewidth]{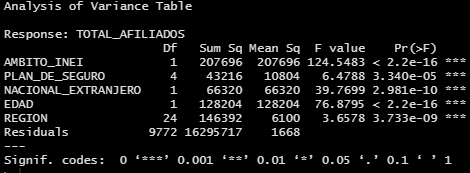}}
        \caption{Result: Validation of the variables}
    \end{figure}

    Interpretation: the p values associated with each variable are very low (less than 2.2e-16), indicating high statistical significance. This suggests that all the variables are important in the model and have a significant effect.\\

    The results indicate that all the variables in the model are statistically significant and have a significant effect on the variability of the dependent variable. These findings support the relevance of considering all the variables included in the model to obtain accurate predictions and a better understanding of the phenomenon studied.\\
    \begin{figure}[htp]
        \centering
        \fbox{\includegraphics[width=.5\linewidth]{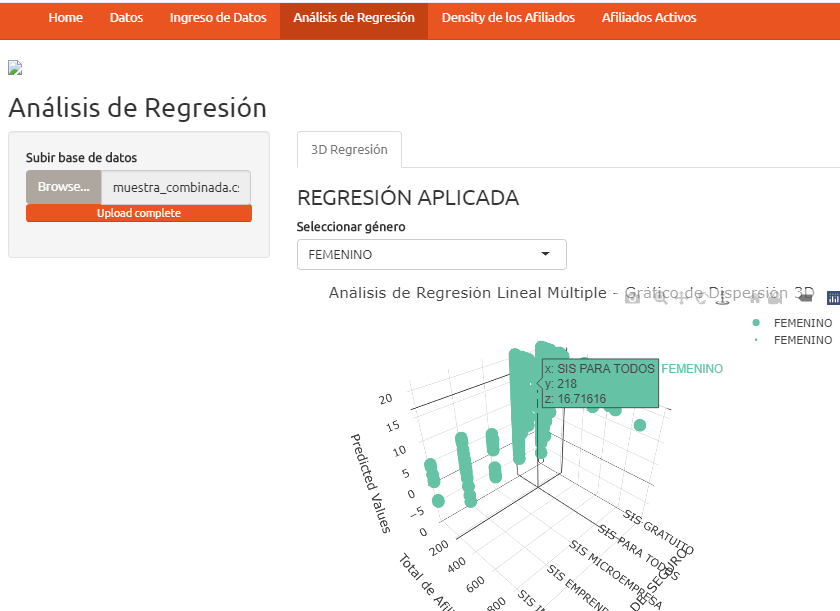}}
        \caption{Result: Multiple Linear Regression}

    \end{figure}
    \\
    The interpretation: when looking at the graph, one can notice a central tendency in the SIS variable for all, which has a positive influence on the total number of affiliates. Furthermore, it can be seen that there is a linear relationship between the variables, indicating that a linear regression model is adequate to describe this relationship.
    In addition, we observed an effective interaction in the application with the user.

    You can view the application at the following url: \\
    
        \url{https://nadineaceituno.shinyapps.io/acti_sis_finesi3d/}

    The database is located at the following url:\\
    
        \url{https://drive.google.com/file/d/1XuHbvOWrYxbI8MxeM2TPT2f6H26Hj_0E/view?usp=sharing}

    Or you can download from the official open data platform at the following url: \\ 
    \url{https://www.datosabiertos.gob.pe/dataset/datos-de-afiliados-al-seguro-integral-de-\\salud-en-estado-activo-seguro-integral-de-salud-13}
\\
\\
        Warning:

        1.Please test before uploading data if you want to extract data from official platform.

        2. In the app, load all the tabs with your database before they are displayed.

        3. In the data tab, refresh the summary before viewing the chart.

        4. In the Affiliate Density tab, press load data before.

\section{DISCUSSION AND CONCLUSIONS}

    In this study entitled "Evaluation of Active Affiliates to the SIS Multivariate Analysis in RShiny", RStudio and RShiny were used as fundamental tools to analyze and understand the factors that affect the affiliation to the Comprehensive Health Insurance.

    Through a systematic approach, a comprehensive exploration of the relevant variables was carried out. The functionalities provided by the application were used for a better outcome, visualization, and observation of the results interactively with the user after the "Home" tab to obtain general information about the SIS, the "Data" tab to load and summarize the affiliate data assets, the "Data Entry" tab to perform additional filters and visualizations, the "Regression Analysis" tab to analyze the relationship between predictor variables and the number of affiliates, the "Affiliate Density" tab to examine the distribution of affiliates, and the "Active Affiliates" tab to view the geographical distribution of affiliates\cite{jia2022development}

    The use of RShiny allowed an interactive exploration of the data, which facilitated the analysis and understanding of the SIS affiliation factors. This provided an intuitive and dynamic visualization of the results, enriching the interpretation and favoring a greater interaction with the data.\cite{alcalde2011sistema}

    Multiple linear regression is a widely used statistical technique for modeling the relationship between a dependent variable and multiple independent or predictor variables.\cite{granados2016modelos} In this article, we will explore the process of fitting a multiple linear regression model and discuss its relevance for predicting the dependent variable.

    The first step in performing multiple linear regression is to use the "lm" function to fit the model. This function takes as its argument the formula that specifies the dependent variable and the independent variables to be used in the model. In the specific case of this article, the dependent variable is the total number of affiliates \textit{(Total Affiliates)}, and the independent or predictor variables include \textit{(Insurance plan, region, age, nationality, foreign, and INEI Scope )} . These variables will be used to predict the dependent variable.

    Once the multiple linear regression model has been fitted using the "lm" function, the result is assigned to a variable called "lm model". This variable contains the estimated coefficients and other information relevant to the model. The estimated coefficients represent the relative contributions of each independent variable to the value of the dependent variable. These coefficients allow us to understand how each independent variable affects the dependent variable and what relative importance each has in the prediction\cite{gomez2016aplicaciones}.\\

    Multiple linear regression has various applications in research and data analysis. It allows analyzing the relationship between multiple variables and predicting the value of a dependent variable based on the independent variables. This technique is used in fields such as economics, sociology, psychology, and data science, among others.\cite{abuin2007regresion}

    Additionally, multiple linear regressions can provide valuable information for decision-making and strategic planning. For example, in the field of insurance, the multiple linear regression model could be used to predict the number of affiliates based on variables such as \textit{(Insurance plan, region, age, national, foreign and INEI scope)}.\cite{hernandez2015determinantes} This information could be used by insurance companies to adjust their marketing strategies, price insurance plans, and make informed decisions about resource allocation.

    These findings contribute to the knowledge and understanding of the elements that influence SIS affiliation and the recurring plan to which they are affiliated, as observed in the multiple linear regression analysis.\cite{dagnino2014regresion}The results obtained through this study provide valuable information for decision-making.

    In conclusion, this study demonstrates the usefulness of RShiny as an efficient tool for the analysis and visualization of data related to comprehensive health insurance affiliation. The results obtained contribute to our knowledge of the factors that influence affiliation and offer a solid base for future research and improvements in health services.
\section{DEDICATION}
    I dedicate this article to all the people who have influenced its realization. First, I would like to thank my parents for their unconditional support, love, and trust in my abilities. This achievement is also thanks to them.

    I am grateful for my career and for all the professors who have guided me and shared their knowledge throughout my academic journey. I valued her dedication and passion for teaching, as well as her inspiration to constantly grow and challenge myself.

    Furthermore, I want to recognize my colleagues and friends, who have been a source of inspiration by sharing moments and giving me mutual support in difficult times. Together, we have built memories and overcome challenges.

    I also want to acknowledge and thank all those who have contributed their work and research in the field of comprehensive health insurance (SIS) and health in Peru. Their contributions have been essential for a better understanding of the factors that affect SIS affiliation and have enriched this article.

    Finally, I thank the readers for their interest in this article. I hope you find our research and analysis useful and contribute to the advancement of knowledge in the area of health and systems with the visual development of RShiny.
\section{Citations}
\label{sec:others}

Multiple linear regression, a highly powered statistical approach, is used to examine the relationship between multiple predictor variables and one response variable. Its application is wide and covers different fields, from economics to data science.
\bigskip
\noindent Multiple Linear Regression. George Seber\cite{abuin2007regresion}.

\bibliographystyle{unsrtnat}
\bibliography{references}  

%%% Uncomment this line and comment out the ``thebibliography'' section below to use the external .bib file (using bibtex) .

%\begin{thebibliography}{1}

%	\bibitem{abuin2007regresion}



% \end{thebibliography}

\end{document}